\documentclass[preprint]{aastex}

\begin{document}

\title{Shock fragmentation model for gravitational collapse}
\author{M. Wilkinson$^{1}$, B. Mehlig$^{2}$ and M. A. Morgan$^{3}$}
\affil{
$^{1}$Department of Mathematics and Statistics, The Open University, Walton Hall, Milton Keynes, MK7 6AA, England\\
$^{2}$Department of Physics, G\"oteborg University, Gothenburg 41296, Sweden\\
$^{3}$Department of Physics, Seattle University, Seattle, WA 98122, USA\\}

\begin{abstract}
A cloud of gas collapsing under gravity will fragment. We present a new theory for this process, in which layers shocked gas fragment due to their gravitational instability. Our model explains why angular momentum does not inhibit the collapse process. The theory predicts that the fragmentation process produces objects which are significantly smaller than most stars, implying that accretion onto the fragments plays an essential role in determining the initial masses of stars. This prediction is also consistent with the hypothesis that planets can be produced by gravitational collapse.
\end{abstract}

\maketitle

%\begin{keywords}
%Gravitational collapse, Jeans mass, Salpeter distribution, shocks,
%fragmentation, opacity.
%\end{keywords}

\section{Introduction}
\label{sec: 1}

Understanding the formation of stars by gravitational collapse is a key problem in astrophysics. Sophisticated simulations (see, for example, \cite{Bat+05,Goo+04}) are used to model star formation within cold molecular clouds in the interstellar medium, but there are significant limitations in the resolution and range of lengthscales achievable with existing computers. Furthermore, it is desirable to have simple models which can be used as a framework for understanding the results of simulations as well as fundamental questions, some of which have not yet been given satisfactory answers.

A theory of star formation should explain why stars have a rather narrow distribution of masses, despite the very broad distribution of physical properties of the molecular clouds. It should also estimate the typical mass of a star and the mass of the smallest objects which are formed by gravitational collapse. When a cloud of gas undergoes gravitational collapse, conservation of angular momentum implies an increase in its angular velocity.  One might expect that this increase in angular speed limits the extent of the collapse, and one would like to estimate the importance of this effect in limiting the collapse process. Also it is known that, on the high-mass side of its peak, the mass distribution is well approximated by a power-law \citep{Sal55}, an observation which theory should explain.

In this paper we propose a model which gives a simplified but physically well-grounded account of gravitational collapse. We show that fragmentation arises because of the gravitational instability of sheets of relatively dense gas produced by shocks. Our theory does not give precise quantitative predictions, but it does give a framework which could be used for producing a quantitative understanding of simulations. We find that angular momentum does not create a barrier to gravitational collapse. We predict that fragmentation ceases when the fragment size is considerably smaller than the mass of a typical star, implying that that mass of a protostar increases by a large factor due to accretion of residual gas after the initial formation of a gravitationally collapsed core (and coalescence of cores could also play a role). There is support from simulations for the hypothesis that accretion is important for determining the initial mass of stars \citep{Bat+05}. A theory for the initial mass distribution must therefore go beyond the distribution of masses of the gravitationally condensed cores which we discuss in this paper. \cite{Bat+05} have described a simple model for this accretion process which gives results consistent with their simulations, but a physically well-motivated model will require substantial additional work.

\section{Other approaches to gravitational fragmentation}
\label{sec: 2}

Before describing our own approach we discuss some of the existing approaches to gravitational collapse. A fundamentally important concept in most discussions of gravitational collapse is the Jeans mass (which is discussed and extended by \cite{Bon57}). A region of initial density $\rho_0$ will collapse by gravitational self attraction in a time of order the Jeans time
\begin{equation}
\label{eq: 2.1}
t_{\rm col}\sim 1/\sqrt{G\rho_0}
\end{equation}
provided gravity can overcome the kinetic energy of relative motion. In an initially near-homogeneous region, Jeans argued that the size of the gravitational collapsing region is determined by the distance that a sound wave, with velocity $c_{\rm s}$, can travel in time $t_{\rm col}$. This leads to the Jeans estimate for the mass produced by gravitational collapse:
\begin{equation}
\label{eq: 2.2}
M_{\rm J}\sim \frac{c_{\rm s}^3}{G^{3/2}\rho^{1/2}}\ .
\end{equation}
The low gas densities which are ascribed to molecular gas clouds give Jeans masses which are usually much larger than the mass of a typical star, indicating that the collapse is accompanied by fragmentation. One theoretical approach to explaining fragmentation stems from the observation that if the collapse proceeds isothermally, the Jeans mass decreases as the density increases, suggesting that the collapsing material may fragment \citep{Hoy53}. It has been argued that the fragmentation will cease when the material is no longer able to cool by emitting radiation as it collapses: in the case of adiabatic collapse the increase of the speed of sound due to increasing temperature means that the Jeans mass ceases to be a decreasing function of density. \cite{Low+76} argued that the collapsing cloud starts to behave adiabatically when it becomes opaque at the Wien wavelength which corresponds to its temperature. They argued that the minimum Jeans mass is insensitive to the opacity of the gas, and used their opacity criterion to estimate the minimum mass for a brown dwarf star, finding a result approximately $7\times 10^{-3}M_\sun$, where $M_\sun$ is the solar mass. The effects of rotation of the collapsing gas are argued to increase this minimum mass, as are magnetohydrodynamic effects where these are relevant.

The arguments put forward by \cite{Low+76} have been used and elaborated by many other authors (for example, \cite{Ree76,Sil77,Mas+99}), but they are subject to two major criticisms. Firstly, it can be questioned whether the Jeans mass is a valid estimate for the size of the fragments, because the Jeans mass concept is only meaningful when the initial density distribution is nearly uniform. This uniformity condition is not satisfied once the gravitational collapse has started. A second criticism concerns the criterion used to determine when the collapsing gas starts to behave adiabatically. \cite{Mas+99} argued that the criterion of the cloud becoming opaque is not relevant, and that the correct approach is to ask when the rate at which thermal energy generated by the gravitational collapse outstrips the ability of the cloud to dissipate radiant heat without an increase in temperature. A further difficulty with the approach in \cite{Low+76} is that it only yields a single estimate for the minimum mass of a star, rather than the probability distribution of masses.

An alternative picture of gravitational collapse, termed \lq turbulent fragmentation' has been proposed \citep{Pad+02,Pad+04}. Simulations of gravitational collapse typically show the production of many supersonic shock waves, which produce a highly non-uniform density of gas. In this approach it is argued that the effects of repeated shocks can produce a log-normal density distribution, with regions where the density is orders of magnitude higher than the typical density of the molecular cloud. This makes it possible to produce regions where the Jeans mass is orders of magnitude smaller than the solar mass, and they use their model to predict the prevalence of brown dwarfs \citep{Pad+04}. At large masses, they advance an argument which relates the observed power-law distribution of masses to the power-law spectrum of turbulent fluctuations via a \lq scale-dependent Mach number', but the dynamics of the collapse process is not incorporated explicitly.  Also, in these models, the critically important parameters of the log-normal distribution are derived from ad-hoc assumptions about the properties of the collapsing gas cloud, and they do not take detailed account of the role of radiative transfer in controlling the collapse of protostars. Our theory also includes the effects of shocks: we give a fuller explanation of the differences between our own theories and those of Padoan and Nordlund in the concluding remarks to our paper (section \ref{sec: 8}). Finally, here we remark that the turbulent fragmentation model builds upon work by \cite{Elm97}, who proposed a theory in which the structure of the collapsing gas cloud is assumed to be a fractal set.

Other discussions of the problem of gravitational collapse emphasise the role of magnetohydrodynamic effects. Magnetic pressure may be very important in stabilising a molecular cloud against collapse \citep{Mes+56}: although the gas is only weakly ionised its effective conductivity is high enough to bind the magnetic field lines, and ambipolar diffusion appears to be too slow to allow disconnection of the field lines \citep{Bas+94}. However we argue that once the collapse is initiated it is a freefall process in which magnetic and hydrostatic pressures play little role until shock waves start to form.

\section{A new approach to collapse and fragmentation}
\label{sec: 3}

Our own theory takes account of the structure of an inhomogeneous collapsing gas cloud by specifying a set of points in a multidimensional parameter space. The structure of the fragmented cloud can be characterised by considering the set of local maxima of gas density at any instant. We assume that gravitational instability will develop each of these density maxima into a discrete fragment. Each of these points may be described by the gas density $\rho$ at that point and also by several other parameters. The most important additional parameter is the size scale, $L$, of the density maximum. If a density maximum is associated with a fragment mass $M$, we define the lengthscale by writing $M=\rho L^3$. If a density maximum is not well separated from its neighbours, we can define its mass by the following procedure. We divide the cloud into Voronoi polygons, based upon the set of positions of the density maxima, and determine the total mass $M_{\rm cell}$ in each of these cells. We define the lengthscale $L$ by writing $M_{\rm cell}=\rho L^3$, where $\rho$ is the maximum density within the cell.

In addition to $\rho$ and $L$, other parameters can be used to characterise the fragments. The most important of these is the angular momentum $J$ of the fragment (relative to its own centre of mass), which can be used to define a characteristic rotation rate $\Omega$ of the cloud fragment through the relation $J=ML^2\Omega$. In addition, thermodynamic variables such as the characteristic temperature and pressure of the fragment could be used, as well as variables describing its magnetohydrodynamic state. However, we are primarily concerned with the behaviour of fragments while they are in isothermal collapse, so that temperature is not a relevant variable. Furthermore, at the level of approximation which is considered below, there is no significant distinction between a gas which dominated by magnetic pressure and one which is dominated by hydrostatic pressure, except that in the former case the speed of sound $c_{\rm s}$ would be replaced by the Alvf\'en speed. We therefore propose to characterise the collapsing gas cloud by a set of points in a three-dimensional space, with coordinates $\rho$, $L$, $\Omega$, representing the peak density, size scale and rotation rate of the region associated with each density maximum. Initially we have just one point, representing the parameters of the entire molecular cloud. As the collapse proceeds, the cloud fragments and the number of representative points increases, until finally each point reaches a region of parameter space where fragmentation no longer occurs. This description could be used to characterise empirically the results of numerical simulations of gravitational collapse, but here we are concerned with a theoretical description of the dynamics and fragmentation of points in $(\rho,L,\Omega)$ space. For simplicity of exposition, in the following discussion we initially neglect the role of angular momentum, giving a description of the dynamics which is confined to the $\Omega=0$ plane. We discuss the effects of including the rotation rate variable in section \ref{sec: 6}, where we show that the evolution of the rotation rate is (approximately) slaved to that of another variable, which characterises the strength of shock waves. In this way we show that the effect of centrifugal effects in resisting collapse does not become more significant as the collapse proceeds.

\begin{figure}
\includegraphics[width=10cm]{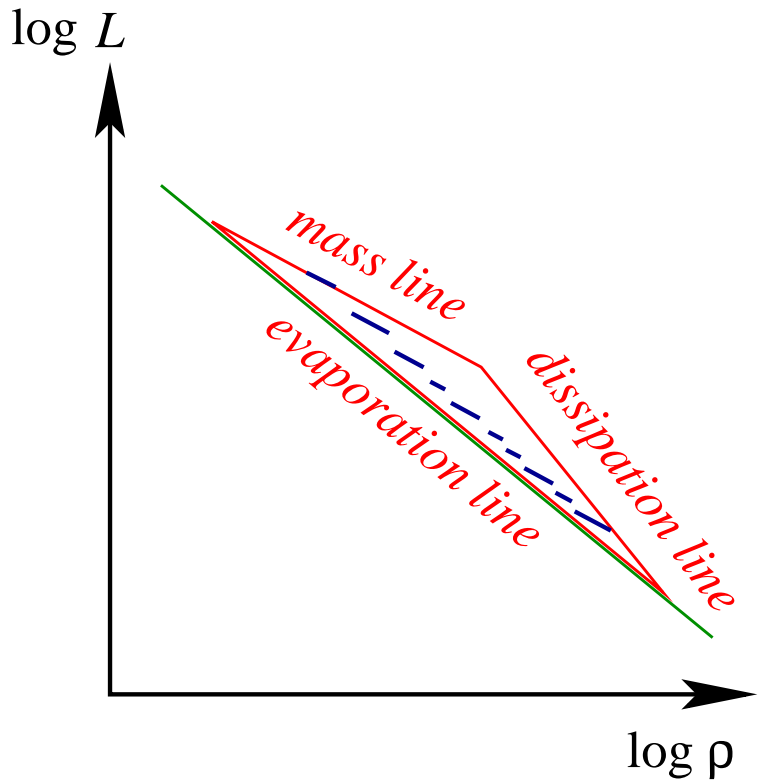}
\caption{\label{fig: 1}
Fragments of a molecular cloud are described by their density $\rho$ and lengthscale $L$. In a logarithmic plot of the parameter space, the fragmenting dynamics is confined to a triangular region. The three lines bounding this region represent the mass of the molecular cloud,  the critical density for stability of fragments against evaporation, and the dissipation limit beyond which the collapse becomes adiabatic, and fragmentation ceases. A typical trajectory has jumps associated with the production of shock waves.}
\end{figure}

It is convenient to use logarithmic coordinates in $(\rho,L)$ space, so we shall discuss the motion of points in the $\log \rho$, $\log L$ plane, illustrated in figure \ref{fig: 1}. As the collapse proceeds, points move downwards (to smaller $L$) and to the right (to higher $\rho$), and a representative point is replaced by two or more points when the density maximum fragments into two or more maxima. If a cloud collapses in a homogeneous free-fall, the evolution is along a line of constant mass, $M=\rho L^3$. Between fragmentation events, the representative points therefore move to the right in $(\log \rho,\log L)$ space, along a straight line of slope $-\frac{1}{3}$, representing evolution at constant mass. When fragmentation occurs the number of representative points increases. We shall argue that the dominant mechanism for this involves the occurrence of shock waves in the collapsing cloud, which implies a discontinuous increase in the density $\rho$, accompanied by a discontinuous reduction in the lengthscale $L$.

We argue that the dynamical processes leading to fragmentation only occur in a certain triangular region of the $(\log \rho, \log L)$ plane. One of the edges of this triangle is a line of slope $-\frac{1}{3}$ which represents a line of constant mass, equal to the original mass of the molecular cloud, $M_0=\rho_0 L_0^3$, and all of the points representing fragments of this cloud must lie below this line. Another constraint is that collapse only proceeds if the escape velocity from the fragment, which is of order $\sqrt{GM/L}=\sqrt{G\rho L^2}$, is large compared to the characteristic molecular velocity, which is of the order of the speed of sound, $c_{\rm s}$. If this condition is not satisfied the fragment can disperse by evaporation. This criterion,
\begin{equation}
\label{eq: 3.1}
\frac{G\rho L^2}{c^2_{\rm s}}\ge K
\end{equation}
(where $K$ is some dimensionless constant of order unity) implies that any fragment which can collapse gravitationally must lie above a line of slope $-\frac{1}{2}$ in the $(\log \rho,\log L)$ plane. The collapse of the gas may be resisted by macroscopic turbulent motion with a typical velocity $v_{\rm turb}$, or by magnetic pressure effects, associated with propagation of disturbances at the Alfv\'en speed, $v_{\rm A}$. If either of these velocities exceeds the speed of sound, we replace $c_{\rm s}$ with the largest of these characteristic speeds in the following calculations. There is another line, which we term the dissipation line, which represents the boundary of the region where the cloud fragments can remain approximately isothermal by radiating away heat. Our own criterion for this is different from that proposed in \cite{Low+76}: we argue that it is not determined by the condition that the cloud has become opaque, and for this reason we use the term dissipation limit rather than opacity limit. We will argue that the collapse remains isothermal provided that $\rho^{3/2}L^2$ remains smaller than some characteristic number. We assume that fragmentation ceases once the isothermal condition is not met. This implies that the representative points must lie below a line of slope $-\frac{3}{4}$ in the $(\log \rho,\log L)$ plane for further fragmentation to be possible. We discuss our dissipation criterion (which is closely related to one proposed by \cite{Mas+99}), and its relation to earlier work, in section \ref{sec: 5}. The fragmentation dynamics is therefore confined to the triangle illustrated in figure \ref{fig: 1}: once the collapse trajectories exit they no longer fragment.
Trajectories which pass the dissipation line continue to collapse. If any trajectories pass the evaporation line, the gas cloud which they represent cannot collapse further: it may expand slowly or else the gas may be accreted by protostars.

To complete the specification of the dynamics, we must consider the process by which representative points undergo fragmentation, as well as describing the dissipation limit. Both of these points require substantial addition discussion. Before embarking on this, we consider the connections between our representation of gravitational collapse and the standard picture based upon the Jeans mass.

In the standard picture of fragmentation during gravitational collapse, as proposed by \cite{Hoy53} and \cite{Low+76}, attention focusses on just one variable, $\rho$, which is used to parametrise the state of the collapsing fragments. The lengthscale is always assumed to be the Jeans lengthscale, $L_{\rm J}=c_{\rm s}t_{\rm col}$, where the Jeans collapse time is $t_{\rm col}=(G\rho)^{-1/2}$, so that the dynamics is assumed to follow a line in the plane illustrated in figure \ref{fig: 1}. On this line the density $\rho$ and lengthscale $L$ satisfy $G\rho L^2\sim c_{\rm s}^2$, so that the non-evaporation condition is only marginally satisfied. In this standard picture, based upon the density dependence of the Jeans mass, the dynamics of the collapse process is therefore envisaged to follow the lower edge of our triangle, with slope $-\frac{1}{2}$, until the dissipation limit is reached and fragmentation stops. The standard approach gives no picture of the mechanism by which fragmentation occurs, nor does it give any framework to estimate the probability distribution of fragment sizes, or the role of angular momentum.

The next section discusses the role of shocks. The dissipation limit for isothermal fragmentation is explained in section \ref{sec: 5}. In section \ref{sec: 6} we discuss the role of angular momentum in limiting the progress of the fragmentation process. In section \ref{sec: 7} we use these results to discuss the probability distribution of fragment sizes. A quantitative theory for the fragment sizes would require information about the frequency with which shocks occur. The relevant statistics would have to be determined by numerical simulation, but it is clear that the distribution of fragment sizes has a \lq long tail' in the high-mass range, which is similar to the Salpeter distribution. Section \ref{sec: 8} summarises our conclusions and discusses the comparison between our theory and \lq turbulent fragmentation'. We argue that the fragment sizes produced by our model are significantly smaller than the sizes of typical stars, so that accretion must be invoked to explain the initial mass function of stars.

\section{The role of shocks}
\label{sec: 4}

In almost all circumstances the collapsing gas cloud fragments will not be spherically symmetric, and the first singular structure to appear as the collapse proceeds will be a shock wave. The time at which the shock wave appears in collpase of a fragment of typical density $\rho$ will be of order $t_{\rm col}=1/\sqrt{G\rho}$, but the exact time at which the shock appears and the its strength will depend upon the initial density and velocity distribution of the cloud. The collapse of a spherical cloud of uniform density is described by a similiarity solution, in which all of the mass concentrates at the centre at the same time, and correspondingly a near-spherical cloud of near uniform density will produce a very strong shock at a late stage of the collapse. However the typical case will result in a shock strength which is of order unity. In the following we will regard the times and strengths of the shock events as random variables with an unknown distribution which we could determine by simulation of an ensemble of collapsing gas clouds.

Pressure plays little role in the free-fall, until the point at which shock waves are generated. There is easily enough energy for freefall to accelerate the gas cloud so that two parts of it will meet at supersonic speed, $v\gg c_{\rm s}$. Where this happens the gas at the contact region is compressed, and shock waves travel out from the compressed region into each of the incoming masses of gas. The shock dissipates most of the relative kinetic energy of the colliding masses of gas into heat. If the relative speed of the collision is $v$, we expect that the thermal molecular velocity in the shocked region is comparable to $v$. Correspondingly, the shocks might be expected to propagate with a speed $V_{\rm s}$ which is some finite fraction of the initial speed $v$. Note that the collision speed may be very large, and correspondingly the temperature of the shocked gas may be very high.

\begin{figure}
\includegraphics[width=9cm]{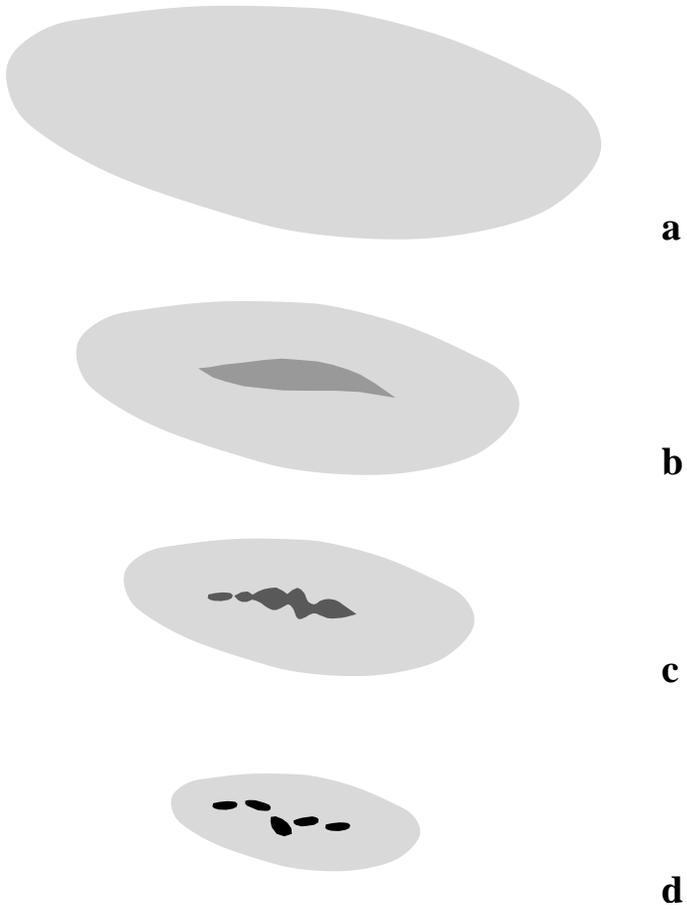}
\caption{\label{fig: 2}
Schematic diagram of a two-dimensional slice through a collapsing fragment. An asymmetric fragment ({\bf a}) collapses in a free-fall until a shock forms ({\bf b}). The shocked layer subsequently puckers due to gravitational instability ({\bf c}) and fragments ({\bf d}). The daughter fragments collapse at a faster rate than their parent.}
\end{figure}

However, it will be argued that the characteristic radiative cooling timescale, $t_{\rm rad}$, is small compared to the Jeans timescale, until the final stages of fragmentation. This indicates that the shocked gas loses energy very quickly by radiating heat, so that actually the shock fronts will move at speeds which are very small compared to $v$. In fact, if the gas cools to the ambient temperature where the speed of sound is $c_{\rm s}$, we expect that the shock front speed $V_{\rm s}$ is comparable to $c_{\rm s}$. The collapsing gas therefore forms sheets which have a very high density, relative to the that of the initial gas cloud: the density is increased by a factor
\begin{equation}
\label{eq: 4.1}
\rho'/\rho = v/V_{\rm s} \sim v/c_{\rm s}\ .
\end{equation}
The shocked material will build up into a sheet of thickness
\begin{equation}
\label{eq: 4.2}
h\sim c_s t_{\rm col}
\end{equation}
where $t_{\rm col}$ is the timescale over which the gas clouds are in collision. The formation of the shock is illustrated in figure \ref{fig: 2}.

If magnetic pressure exceeds the hydrostatic pressure, we have to replace $c_{\rm s}$ by the higher speed of magnetosonic waves (the Alfv\'en speed), but the basic principle remains unchanged. That is, energy is provided by freefall and efficiently dissipated to an isothermal state by a shock which cools rapidly.

Consider the collapse of a cloud fragment of size $L_0$ and density $\rho_0$. As the collapse proceeds, freefall will take these values to $\rho$, $L$ along a line of constant mass $M=\rho L^3$. After the fragment has collapsed to some fraction $f=L/L_0$ of its original size, a shock wave starts to form. The value of $f$ depends on the details of the initial phase-space distribution of the mass within the fragment. When the lengthscale is $L$ we expect that the collision velocity of colliding gas clouds is
\begin{equation}
\label{eq: 4.3}
v \sim \sqrt{GM/L}=\sqrt{G\rho L^2}\ .
\end{equation}
The characteristic time over which the shock wave propagates through the gas cloud is the same as the characteristic time for gravitational collapse of the fragment, namely $t_{\rm col} \sim 1/\sqrt{G\rho}$.
Thus, the shocks generated by collision of masses of gas when the size
of the fragment is $L$ form sheets with an area of order $L^2$ and thickness of order
\begin{equation}
\label{eq: 4.4}
L'\sim h\sim t_{\rm col} c_{\rm s}\sim \frac{c_{\rm s}}{\sqrt{G\rho}}\ .
\end{equation}
The shock therefore creates a layer of material of thickness $L'$ and density
\begin{equation}
\label{eq: 4.5}
\rho'=\rho\frac{L}{L'}=\frac{G^{1/2}\rho^{3/2}L}{c_s}\ .
\end{equation}
It will be instructive to express $\rho'/\rho$ and $L'/L$ in terms of the ratio $M/M_{\rm J}$, where $M=\rho L^3$ and $M_{\rm J}=c^3_{\rm s}/G^{3/2}\rho^{1/2}$ is the Jeans mass corresponding to $\rho$. We find:
\begin{equation}
\label{eq: 4.6}
\frac{\rho'}{\rho}\sim \frac{v}{c_{\rm s}} =\sqrt{\frac{G\rho L^2}{c_{\rm s}^2}}=\frac{G^{1/2}\rho^{1/6}(\rho L^3)^{1/3}}{c_{\rm s}}=\left(\frac{M}{M_{\rm J}}\right)^{1/3}
\end{equation}
and
\begin{equation}
\label{eq: 4.7}
\frac{L'}{L}\sim \frac{c_{\rm s}}{\sqrt{G\rho L^2}}=\left(\frac{M_{\rm J}}{M}\right)^{1/3}\ .
\end{equation}
Having formed sheets of relatively dense gas, with surface density $\Sigma\sim \rho' L'$, these will now themselves undergo gravitational collapse. The gravitational instability of a sheet of gas with surface density $\Sigma$ which rotates at a uniform angular velocity $\Omega$ is a problem which has been solved (see, for example, \cite{Bin+87}), and we base our discussion upon its solution. The sheets of gas produced by the shock waves will in general be a curved surface, with varying thickness, density and in-plane velocity gradient (which may include shear as well as rotation), but the results for a uniformly rotating flat sheet contain the essential physics. The collapse occurs on a timescale $t'_{\rm col}=\gamma^{-1}$, where the instability rate $\gamma$ describing exponential growth of a density perturbation with wavenumber $k$ satisfies
\begin{equation}
\label{eq: 4.8}
\gamma^2=2\pi G \Sigma \vert k\vert - v_{\rm s}^2k^2 -4\Omega^2
\end{equation}
where $v_{\rm s}$ is a sound speed characterising the propagation of sound waves in the sheet and $\Omega$ is the local angular velocity of the material in the sheet, about an axis perpendicular to the surface \citep{Bin+87}. When this expression yields a negative value for $\gamma^2$, the sheet is stable.

Consider the interpretation of (\ref{eq: 4.8}) for the sheet of shocked material in the case where $\Omega=0$. For disturbances of the sheet which have a wavelength $2\pi /k$ which is comparable to or less than the sheet thickness, it is necessary to consider the stability problem in three dimensions (so that (\ref{eq: 4.8}) ceases to be applicable). The largest allowed value of $k$ is therefore of order $2\pi/L'$, where $L'$ is the thickness of the sheet. The inequality (\ref{eq: 3.1}) then implies that the second term is negligible for fragments which are far away from the evaporation line in figure \ref{fig: 1}. Equation (\ref{eq: 4.8}) then implies fragmentation: the larger $k$, that is the smaller the fragments, the more rapid the collapse, indicating that the sheet fragments into pieces of size of order $L'$. This fragmentation has a characteristic timescale $t_{\rm frag}\sim 1/\sqrt{G\rho'}$, which is faster than the timescale for the collapse of the parent fragment by a ratio $\sqrt{v/c_{\rm s}}$. When $\Omega $ is sufficiently large, however, the sheet is stabilised against fragmentation: we return to consider this in detail in section \ref{sec: 6}.

Having created smaller fragments of size $L'$ and density $\rho'$, these fragments will themselves collapse by freefall, and produce additional shocks. Let us consider the how the shock and fragmentation process is represented on the diagram of the parameter space in figure \ref{fig: 1}. We parametrise the fragments by their density and by their distance from the Jeans line using coordinates
\begin{equation}
\label{eq: 4.9}
X=\log \rho\ ,\ \ \ Y=\log (M/M_{\rm J})
\end{equation}
respectively. The shock produces jumps in the values of $X$ and $Y$. From (\ref{eq: 4.6}), we see that the jump in $X$ is equal to
$\Delta X=\frac{1}{3}Y$. After the shock, from (\ref{eq: 4.5}) and (\ref{eq: 4.7}) we find that the new value of $Y$ is
\begin{eqnarray}
\label{eq: 4.10}
Y'&=&\log \left(\frac{M'}{M'_{\rm J}}\right)=\log\left(\frac{\rho'L'^3G^{3/2}\rho'^{1/2}}{c_{\rm s}^3}\right)
\nonumber \\
&=&\log\left(\rho L^3\frac{G^{3/2}\rho^{1/2}}{c_{\rm s}^3}\right)-\frac{1}{2}Y=\frac{1}{2}Y\ .
\end{eqnarray}
When a shock forms the values of the coordinates $X$ and $Y$ therefore jump to
\begin{equation}
\label{eq: 4.11}
X'=X+\frac{1}{3}Y\ ,\ \ \ Y'=\frac{1}{2}Y\ .
\end{equation}
At each shock the number of density maxima (and correspondingly the number of representative points) is multiplied by a factor $N$. The largest possible value of $N$ is of order $(L/L')^2=\exp(2Y/3)$, but the value may be much smaller. In particular, if the rotation rate of the sheet is sufficiently large, fragmentation may not be possible at all. Between shocks, the representative points move along a line of constant slope, representing freefall at constant mass,
\begin{equation}
\label{eq: 4.12}
M=\rho L^3=\frac{M}{M_{\rm J}}\rho^{-1/2}\frac{c_{\rm s}^3}{G^{3/2}}
\end{equation}
so that the freefall line has slope $\frac{1}{2}$ in the $(X,Y)$ parametrisation.

\begin{figure}
\includegraphics[width=10cm]{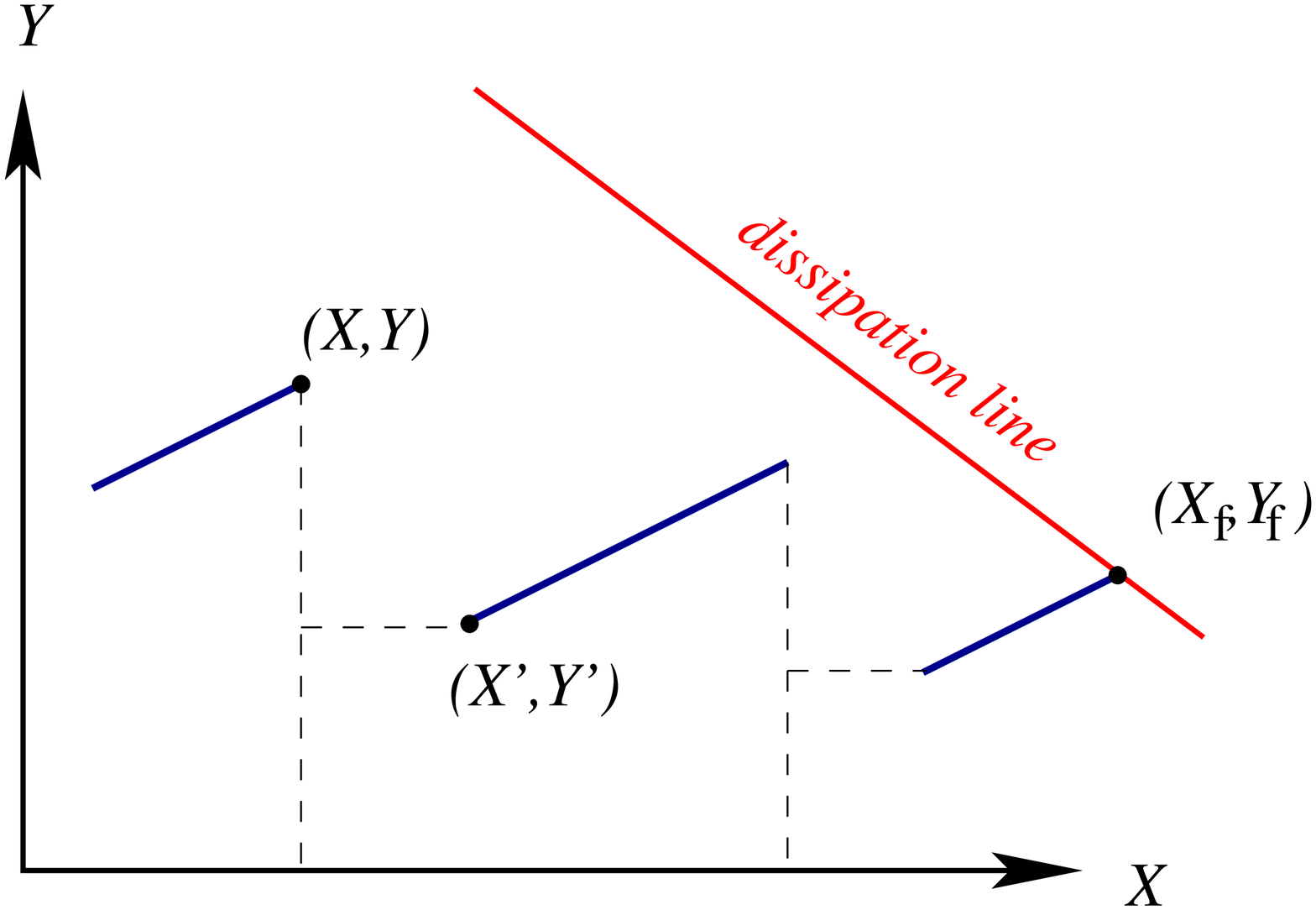}
\caption{\label{fig: 3}
Schematic illustration of the motion of a representative point in the coordinates $X=\log \rho$, $Y=\log (M/M_{\rm J})$. The blue lines are intervals of free-fall collapse, which terminate with formation of a shock. The shocked surface forms fragments which have new coordinates determined by (\ref{eq: 4.11}). Fragmentation ceases when the trajectory meets the dissipation line at $(X_{\rm f},Y_{\rm f})$, which in these coordinates has slope $-\frac{3}{4}$.}
\end{figure}

To summarise, the dynamics of the collapse process involves a series of jumps in the $(\log \rho,\log L)$ plane, which are associated with the formation of shocks in collapsing fragments. A typical trajectory is shown schematically in figure \ref{fig: 1}. Between the shocks, the representative points evolve along lines of constant mass, which take them further from the evaporative stability line, where $M=M_{\rm J}$. The shocks create new fragments closer to this line. The collapse process may be understood by following the trajectories of representative points in the $(X,Y)$ plane, with coordinates defined by equation (\ref{eq: 4.9}). The motion of the representative points in these coordinates is illustrated schematically in figure \ref{fig: 3}.

We emphasise that the foregoing is not to be interpreted as a precise description of the very complicated dynamics of the gravitational collapse process, but rather a model indicating the order of magnitude of the effects of the shock waves.

\section{Condition for gas to remain isothermal}
\label{sec: 5}

The large density enhancement caused by the shock waves discussed in section \ref{sec: 4} depends upon the shocked material radiating heat so rapidly that it remains approximately isothermal. If the shock waves become adiabatic, the density enhancement is or order unity, and shocks become much less effective at seeding fragmentation. We assume that fragmentation ceases when the shocks become adiabatic. Various criteria for the collapsing gas to become adiabatic have been discussed \citep{Low+76,Ree76,Mas+99}. Our own approach is very close to that of \cite{Mas+99}, but not close enough that we can simply quote their results. We discuss how our arguments compare with those in earlier works at the end of this section.

First we argue that the timescale $t_{\rm rad}$ for an element of gas to equilibrate with its surrounding by radiative transport is very short compared to the timescale for gravitational collapse, justifying the idea that isothermal collapse is possible. A small element of gas, with volume $\Delta V$ which is at temperature $T$ and mass density $\rho$ contains internal energy
\begin{equation}
\label{eq: 5.1}
\Delta U=\frac{3\rho kT}{2m_0}\Delta V
\end{equation}
where $m_0$ is the molecular mass, and $k$ is Boltzmann's constant, and we assume that the temperature is sufficiently low that the hydrogen molecular rotation spectrum is not excited. This element radiates heat to its surroundings at a rate
\begin{equation}
\label{eq: 5.2}
\dot Q=4\kappa \rho \sigma T^4 \Delta V
\end{equation}
where $\sigma$ is the Stefan-Boltzmann constant and $\kappa$ is the Planck mean opacity. We see that the time constant for the radiative transfer of energy, $t_{\rm rad}=\Delta U/\dot Q$, satisfies
\begin{equation}
\label{eq: 5.3}
\frac{1}{t_{\rm rad}}=\frac{8\kappa m_0\sigma T^3}{3 k}\ .
\end{equation}
In the molecular clouds of the interstellar medium, the optical properties at the Wien wavelength for black-body radiation at the temperature of the molecular cloud (approximately $10\,{\rm K}$) are dominated by dust particles, and scattering is insignificant relative to absorption. The opacity at $T=10\,{\rm K}$ is estimated (see, for example \cite{Bec+90}) to be approximately
\begin{equation}
\label{eq: 5.4}
\kappa\approx 10^{-3}\,{\rm m}^2{\rm kg}^{-1}\ .
\end{equation}
The metallicity of stars varies by about an order of magnitude, indicating that there is at least an order of magnitude uncertainty in the value of the opacity. The isothermal speed of sound of interstellar gas at $10\,{\rm K}$ is $c_{\rm s}\approx 190\,{\rm ms}^{-1}$ (assuming a mean molecular mass of $2.34$). Using these values we estimate $t_{\rm rad}\approx 2.3\times 10^{10}\,{\rm s}\approx 750\,{\rm yrs}$. This timescale is short enough to ensure that radiative transfer maintains isothermal conditions until the late stages of the collapse process.

The rate at which heat is generated in a collapsing cloud is of order
\begin{equation}
\label{eq: 5.5}
\dot Q_{\rm gen}\sim \frac{GM^2}{Lt_{\rm col}}\sim G^{3/2}\rho^{5/2}L^5\ .
\end{equation}
The collapse ceases to be isothermal when this rate of heat production exceeds the rate at which the cloud fragment can radiate heat. This latter question depends upon whether the cloud is transparent or opaque at the Wien wavelength corresponding to the cloud temperature.

It turns out that the optically thin case is most relevant, and we consider this first. For an optically thin cloud of volume $V=L^3$ which is at a temperature $T+\Delta T$ and which is surrounded by gas at temperature $T$, the rate at which heat is radiated is
\begin{equation}
\label{eq: 5.6}
\dot Q_{\rm thin}=4\kappa \rho \sigma [(T+\Delta T)^4-T^4] L^3
\end{equation}
where $\kappa$ is the opacity, provided that the optical thickness is small, $\kappa\rho L\ll 1$. The collapse ceases to be isothermal when $\Delta T/T$ is of order unity. Combining (\ref{eq: 5.5}) and (\ref{eq: 5.6}), the condition for near-isothermal collapse to cease is
\begin{equation}
\label{eq: 5.7}
\rho^{3/2}L^2\sim \frac{\kappa \sigma T^4}{G^{3/2}}
\end{equation}
provided $\kappa \rho L\ll 1$. This condition corresponds to a line of slope $-3/4$ in the $(\log L,\log \rho)$ plane, or equivalently to a line in the $(X,Y)$ plane (defined by (\ref{eq: 4.9})) which also has slope $-3/4$.

Now consider the smallest possible fragment mass in the case where the dissipation limit is determined by (\ref{eq: 5.7}), which corresponds to the point at which the dissipation line intersects the evaporation line, specified by (\ref{eq: 3.1}). Combining (\ref{eq: 3.1}) and (\ref{eq: 5.7}) we find that the density at the intersection point is
\begin{equation}
\label{eq: 5.8}
\rho_{\rm max} \sim \frac{\left({\kappa \sigma T^4}\right)^2}{Gc_{\rm s}^4}\ .
\end{equation}
It is instructive to note that this estimate is the same as that obtained by equating the gravitational collapse time, $t_{\rm col}=(G\rho)^{-1/2}$ to the radiative transfer time given by (\ref{eq: 5.3}). Using the estimate (\ref{eq: 5.3}) in equations (\ref{eq: 5.8}) and (\ref{eq: 2.2}), we find the following estimates for the maximum density at which fragmentation occurs, $\rho_{\rm max}$, and the corresponding fragment mass, $M_{\rm min}$:
\begin{equation}
\label{eq: 5.9}
\rho_{\rm max}\approx 3.9\times 10^{-12}\, {\rm kg}\,{\rm m}^{-3}\ \ \ \
M_{\rm min}\approx 6\times 10^{27}\,{\rm kg}\ .
\end{equation}
With these data the assumption that the gas has not become opaque is satisfied: $\kappa \rho L\sim \kappa \rho^{1/2} c_{\rm s}G^{-1/2}\approx 0.05$.

The minimum mass is small compared to that of a star, by about two orders of magnitude. Even accounting for uncertainty in the dimensionless prefactors and that most fragments will be significantly heavier than the minimum mass, this estimate indicates that the smallest fragments produced in gravitational collapse are not protostars, but rather cores upon which material can accrete to become protostars. The minimum mass is, however, comparable to that of a gas-giant planet, which makes it plausible that gas-giant planets with eccentric orbits could have been formed by gravitational collapse, rather than by the aggregation of dust grains \citep{Rib+07,Wil+08}.

The opacity may differ from the estimate (\ref{eq: 5.3}), and studies of stellar atmospheres indicate that the metallicity of stars varies over approximately an order of magnitude. If the opacity $\kappa$ is multiplied by a factor $\theta$, then $\rho_{\rm max}$ is multiplied by $\theta^2$ and the minimum mass of a gravitationally condensed core is multiplied by $\theta^{-1}$: that is, higher amounts of dust will favour smaller cores.

When the gas becomes optically thick, the transport of heat by radiation is then determined by diffusion, and heat may be able to escape from the fragment at a sufficient rate to allow it to remain isothermal. We found above that for the value of opacity quoted in (\ref{eq: 5.3}) the dissipation line is determined by the optically thin case, but that the extinction length is only marginally larger than the cloud fragment. We should therefore also discuss the optically thick case, where diffusive transport of radiation is the important mechanism.

There are two types of radiative diffusion behaviour, depending upon whether the thermal radiation is predominantly scattered or absorbed and re-emitted. In the scattering case the radiative diffusion constant is $D_{\rm rad}\sim \Lambda c$, where $\Lambda $ is the scattering mean free path, and $c$ is the speed of light. However, if absorption dominates the opacity, the diffusion of radiative energy is characterised by a diffusion coefficient
\begin{equation}
\label{eq: 5.10}
D_{\rm rad}\sim \frac{1}{\alpha^2 t_{\rm rad}}
\end{equation}
where $t_{\rm rad}$ is the timescale for emission of black-body radiation from the absorbing elements, and $\alpha=\kappa\rho$ is the inverse extinction length.
Using (\ref{eq: 5.3}), we find the diffusion coefficient
\begin{equation}
\label{eq: 5.11}
D_{\rm rad}\sim \frac{m_0\sigma T^3}{\kappa \rho^2 k}\ .
\end{equation}
If the temperature varies slowly on the scale of the optical depth, it satisfies a diffusion equation and the radiative heat flux is
\begin{equation}
\label{eq: 5.12}
\mbox{\boldmath$q$}=\frac{k\rho }{m_0}D_{\rm rad}\mbox{\boldmath$\nabla$}T
\ .
\end{equation}
The the rate of heat production $\dot Q_{\rm gen}$ in a collapsing fragment of size $L$ is of order $GM^2/Lt_{\rm col}$, where $M=\rho L^3$ is the mass of the fragment and $t_{\rm col}=(G\rho)^{-1/2}$ is its characteristic collapse time. The interior of the fragment will exceed the ambient temperature by an amount $\Delta T$. Estimating the heat flux as $\vert \mbox{\boldmath$q$}\vert \sim \dot Q/L^2$, we estimate $\Delta T\sim \dot Q m_0/\rho kLD_{\rm rad}$. The condition for approximate isothermality, $\Delta T/T\ll 1$, can then be expressed in the form
\begin{eqnarray}
\label{eq: 5.13}
\frac{\Delta T}{T}&\sim &\frac{\dot Q}{L}\frac{m_0}{\rho kTD_{\rm rad}}\sim \frac{L^2}{D_{\rm rad}t_{\rm col}}\frac{G\rho L^2}{c_{\rm s}^2}
\nonumber \\
&\sim & \frac{L^2}{D_{\rm rad}t_{\rm col}}\left(\frac{v}{c_{\rm s}}\right)^2\ll 1\ .
\end{eqnarray}
where $v/c_{\rm s}$ is the shock strength parameter introduced in section \ref{sec: 4}. From (\ref{eq: 5.7}), we find that the condition $\Delta T/T=O(1)$ is satisfied when $L^4\rho^{7/2}$ is equal to some constant. We conclude that if the dissipation line is determined by the diffusive condition, then the dissipation line in figure \ref{fig: 1} has slope $-\frac{7}{8}$, instead of $-\frac{3}{4}$.

Finally we comment on some earlier works which have proposed different criteria for the collapse to become adiabatic. \cite{Low+76} argued that the collapse becomes adiabatic when the gas becomes opaque on the scale of the Jeans length. This criterion is most widely referred to, but it does not appear to be supported by any calculation. These authors also propose that in the period before the collapse becomes adiabatic, the temperature of the gas rises as $T\propto \rho^{1/6}$, rather than remaining isothermal, which seems to be due an incorrect application of the energy balance relation for a volume element. \cite{Ree76} gives a discussion of what is essentially our own criterion for the dissipation line, but his paper describes the results as if they are consistent with the opacity criterion in \cite{Low+76}. \cite{Mas+99} were the first to point out that the opacity condition for adiabaticity is not correct, and they appear to be the first to point out that diffusion may limit the rate of radiation transfer when the gas cloud is opaque. Their discussion assumes implicitly that the collapse process is parametrised by a single variable (the gas density). Our own results extend theirs by describing the dissipation limit as a line in the $(\log \rho,\log L)$ plane.

\section{The role of angular momentum}
\label{sec: 6}

If a cloud of gas is rotating, the centrifugal effect resists gravitational collapse. The cloud can only collapse if the gravitational collapse rate $\sqrt{G\rho}$ is sufficiently large relative to the angular velocity of the cloud, $\Omega$. Thus collapse only proceeds if
\begin{equation}
\label{eq: 6.1}
\frac{G\rho}{\Omega^2}\ge K'
\end{equation}
where $K'$ is a dimensionless constant of order unity.

The angular momentum $J$ of an isolated gas cloud is conserved, suggesting that as the cloud collapses the rotation rate $\Omega$ will increase and eventually the inequality (\ref{eq: 6.1}) will not be satisfied, so that gravitational collapse will cease. If the size scale of the cloud is $L$ and the density is $\rho$, then the angular momentum $J$ is of order $J\sim ML^2\Omega$, where $M\sim \rho L^3$ is the mass of the cloud. In a freefall collapse of the cloud, the conservation of angular momentum implies that $\Omega\propto L^{-2}$, and the conservation of mass implies $L\propto \rho^{-1/3}$, so that $\Omega \propto \rho^{2/3}$. It follows that $G\rho/\Omega^2\propto \rho^{-1/3}$, implying that the freefall collapse of a gas cloud will be halted when the density $\rho$ becomes sufficiently large.

If the gas cloud undergoes fragmentation, however, centrifugal effects can become less of a barrier to the collapse proceeding. Note that when we characterise a fragment of a gas cloud, we should be concerned with its angular momentum about its own centre of mass. Thus if a cloud with angular momentum $J$ breaks up into fragments with angular momenta $J_i$, the principle of conservation of angular momentum does not imply that $\sum_iJ_i=J$. This is because each of the $J_i$ is measured relative to an origin which is different from that used to determine $J$, so that in fact we have $J>\sum_i J_i$.

When a cloud fragments by gravitational self-attraction, we expect that as the fragments start to become distinct from the original cloud they will have approximately the same angular velocity as that part of the cloud from which they form. Although there is no exact conservation principle which is applicable, we can assume that the rotation rate of each fragment, $\Omega_i$, is comparable to the rotation rate of the parent cloud: $\Omega_i\sim \Omega$.

The fragmentation mechanism described in section \ref{sec: 4}, which involves the production of strong shocks, produces fragments which have a greater density than the parent cloud, but the rotation rate of the gas will not be significantly changed by the formation of a shock. Equation (\ref{eq: 6.1}) implies that the increased density favours gravitational collapse.

There are thus two competing effects. A freefall collapse inhibits gravitational collapse by increasing the centrifugal effect, whereas the shock-induced fragmentation mechanism favours gravitational collapse by increasing the density. We now consider how to quantify the balance between these opposing effects. To this end we introduce a further logarithmic variable characterising the state of a gas cloud or one of its fragments:
\begin{equation}
\label{eq: 6.2}
Z=\log\left(\frac{G\rho}{\Omega^2}\right)\ .
\end{equation}
The cloud can collapse if $Z$ exceeds a threshold value, which is of order unity. We argued that in freefall, $G\rho\
/\Omega^2\propto \rho^{-1/3}$, so that in the freefall phases the relation between $Z$ and $X$ (defined by (\ref{eq: 4.9})) is $Z=Z_0-\frac{1}{3}X$ (where $Z_0$ is a constant determined by the initial conditions). In freefall, we also have the relation $Y=Y_0+\frac{1}{2}X$ (see equation (\ref{eq: 4.12}); $Y_0$ is another constant), so that if a cloud undergoes freefall such that $X$ changes by $\Delta X$, the corresponding changes in $Y$ and $Z$ are related by
\begin{equation}
\label{eq: 6.3}
\Delta Z=-\frac{2}{3}\Delta Y=-\frac{1}{3}\Delta X\ .
\end{equation}
When the shock and fragmentation process described in section \ref{sec: 4} occurs, however, the density increases with $\Omega$ remaining constant, resulting in an equal jump of both $X$ and $Z$, specified by equation (\ref{eq: 4.11}), so that the value of $Z$ after the shock is
\begin{equation}
\label{eq: 6.4}
Z'=Z+\frac{1}{3}Y\ .
\end{equation}
Equations (\ref{eq: 6.3}) and (\ref{eq: 6.4}) show that the evolution of the variable $Z$, which describes the effect of centrifugal forces, is slaved to the evolution of $Y$, which parametrises the strength of shocks.
In the shock, equation (\ref{eq: 4.11}) also shows that the value of $Y$ is halved. Thus we see that the evolution of $Z$ is coupled to that of $Y$.
During periods of freefall, $X$ changes by increments $\Delta X_i$, and there is a corresponding change $\Delta Y_i=\frac{1}{2}\Delta X_i$ in the value of $Y$. Let the values of $Y$ just before shock events be $Y_i$. The change in value of $Y$ over many shocks and freefalls is
\begin{equation}
\label{eq: 6.5}
Y(X_{\rm f})-Y(X_{\rm i})\sim \sum_i \frac{1}{2}\Delta X_i-\frac{1}{2}Y_i
\ .
\end{equation}
The value of $Y$ is expected to reach a stationary distribution, in which case $Y(X_{\rm f})-Y(X_{\rm i})$ remains bounded, so that (\ref{eq: 6.5}) implies that
\begin{equation}
\label{eq: 6.6}
\langle \Delta X_i \rangle = \langle Y_i \rangle \ .
\end{equation}
By analogy with (\ref{eq: 6.5}), the change in $Z$ is
\begin{equation}
\label{eq: 6.7}
Z(X_{\rm f})-Z(X_{\rm i})\sim \sum_i -\frac{1}{3}\Delta X_i +\frac{1}{3}Y_i
\end{equation}
Using (\ref{eq: 6.6}), we conclude that the difference between the final and initial values of $Z$ has no drift. This is consistent with $Z$ also approaching a stationary distribution.

Although centrifugal effects may stop some fragments from undergoing further fragmentation, our results are consistent with the hypothesis that angular momentum does not become more significant as the collapse process progresses.

\section{Statistical model for masses}
\label{sec: 7}

An advantage of considering the dynamics of the collapse process in greater detail is that it gives a framework within which the distribution of fragment masses can be predicted. A quantitative calculation of the fragment mass distribution from our model is a difficult problem which will require substantial additional effort. However it is of interest to consider how this calculation can proceed in principle. In the following we discuss in outline how the fragment mass distribution could be calculated. Our discussion indicates that in the limit where the size of the initial molecular cloud becomes very large, the distribution of fragment masses becomes universal. We also argue that our approach is consistent with a power-law for the large-mass tail of the distribution, analogous to the Salpeter distribution (although the typical fragment mass is much smaller than the typical initial stellar mass).

The probability distribution of masses of the fragmentation products is determined by the density of points in the $(X,Y)$ plane at the point where the fragmentation ceases, that is when the representative point passes through the dissipation line. Given this density, $w(X,Y)$, we can determine the probability distribution $P_{\rm f}(Y_{\rm f})$ for the final value of $Y$, that is the value $Y_{\rm f}$ where the fragment trajectory passes the dissipation line. This distribution can then be transformed to determine the final fragment masses, $M=M_{\rm J}\exp(Y)$. We first consider the form of the density $w(X,Y)$, and then discuss how this can be used to determine the fragment mass distribution.

We can define a stochastic process to model the variable $Y(X)$. We consider a process where $Y(X)$ grows linearly with $X$ (with gradient $\epsilon$, say) until a shock occurs, at coordinate $X_i$. Let the probability of an interval of growth between shocks being greater than $\Delta X$ be $P(\Delta X)$. When a shock occurs, the value of $Y$ is multiplied by a fraction $\mu$. We have seen that $\epsilon=\frac{1}{2}$ and $\mu=\frac{1}{2}$ are the appropriate values to use with our model. If the molecular cloud is orders of magnitude more massive than a single star, there will be many fragmentation events during the gravitational collapse, and in this case the steady-state probability distribution of our stochastic process will approach a limit, with probability density function $p(Y)$. The number of fragments is expected to increase exponentially as the fragmentation proceeds, so that the density of fragments in the $(X,Y)$ plane which results from fragmentation of a very large body may be approximated by
\begin{equation}
\label{eq: 7.1}
w(X,Y)=\exp(\eta X)p(Y)
\end{equation}
where $\eta$ is a universal constant and $p$ is a universal function.

Now consider how the fragment mass distribution may be obtained from (\ref{eq: 7.1}). We argued that the dissipation limit is determined by the relation $\rho^{3/2}L^2={\rm const.}$, which corresponds to a line of slope $-\frac{3}{4}$ in the $(X,Y)$ plane, with coordinates defined by equation (\ref{eq: 4.9}). Fragmentation stops when the representative point describing a fragment crosses the dissipation line, at the point $(X_{\rm f},Y_{\rm f})$. We want to compare the mass $M_{\rm f}$ of this fragment with that of the minimum mass fragment, $M_{\rm min}$, corresponding to the rightmost apex of the triangle in figure \ref{fig: 1}, at coordinates $(X_{\rm max},0)$. Because the slope of the dissipation line in the $(X,Y)$ coordinates in $-\frac{3}{4}$, we have $X_{\rm f}=X_{\rm max}-\frac{4}{3}Y_{\rm f}$. Noting that the Jeans mass $M_{\rm J}$ is proportional to $\rho^{-1/2}$, we have
\begin{eqnarray}
\label{eq: 7.2}
\frac{M_{\rm f}}{M_{\rm min}}&=&\frac{M_{\rm J}}{M_{\rm min}}\exp(Y_{\rm f})
=\exp\left(\frac{X_{\rm max}-X_{\rm f}}{2}\right)\exp(Y_{\rm f})
\nonumber \\
&=&\exp\bigg(\frac{5}{3}Y_{\rm f}\bigg)\ .
\end{eqnarray}
To determine the distribution of fragmentation product masses we must therefore evaluate the distribution of the coordinate $Y_{\rm f}$ and use equation (\ref{eq: 7.2}) to relate $Y_{\rm f}$ to the fragment mass. The density at the position $(X_{\rm f},Y_{\rm f})$ relative to that at the corner of the dynamical triangle is:
\begin{equation}
\label{eq: 7.3}
\frac{w(X_{\rm f},Y_{\rm f})}{w(X_{\rm max},0)}
=\exp\left(-\frac{4}{3}\eta Y_{\rm f}\right)\frac{p(Y_{\rm f})}{p(0)}\ .
\end{equation}

It is a difficult task to determine the probability $P(\Delta X)$. In the following we consider the result of applying the simplest model, which is to assume that the probability of going from $X$ to $X+\delta X$ without the formation of a shock is independent of $X$ (that is, we assume that $\Delta X$ has a Poisson distribution). In this case $P(\Delta X)=\exp(-\lambda \Delta X)$, for some constant $\lambda$. We are really interested in the distribution for the case where $\mu=\frac{1}{2}$, but it is difficult to achieve a solution in that case. Rather, we discuss the behaviour of the solution in the limit as $\mu\to 0$, which is very easily obtained (as we show in the following paragraph). Numerical studies show that when $P(\Delta X)$ is Poisson distributed, the tail of the distribution is very similar when $\mu =\frac{1}{2}$.

In the limit where $\mu=0$, the coordinate $Y$ is reduced to zero every time a shock occurs. In this case the probability that $Y$ is less than $Y_0$ is $1-P(Y_0/\epsilon)$. Let us consider the distribution of fragment masses in the case where $P(\Delta X)$ is approximated by a Poisson distribution, so that the probability density of $Y$ is then $p(Y)=A\exp(-\lambda Y/\epsilon)$ (where $A$ is a normalisation constant). If $Y$ has this exponential distribution, the corresponding distribution for $M$ is a power law. Recalling that the fragment mass $M_{\rm f}$ is related to $Y$ by equation (\ref{eq: 7.2}), the probability distribution of fragment masses is of the form
\begin{equation}
\label{eq: 7.4}
p_M(M_{\rm f})\sim C\left(\frac{M_{\rm f}}{M_{\rm min}}\right)^{-({\frac{6\lambda}{5}+\frac{4\eta}{5}+1})}
\end{equation}
in the large-mass limit, where $C$ is a normalisation constant (note that we have set $\epsilon=\frac{1}{2}$ here). The distribution of separations between random events is often characterised by a Poisson distribution, and we have seen that such a distribution gives a power-law tail of the mass distribution, which is analogous to the Salpeter distribution for the masses of stars. However, we observed that the masses of the fragments produced by gravitational collapse are much smaller than the masses of protostars. The initial accretion of mass onto the fragments generated by gravitational collapse is not well understood, and it is uncertain how or whether the distribution of initial masses is reflected in the final masses of protostars. Numerical studies of the collapse process are required to identify the value of $\eta$ and to determine how well $p(Y)$ is modelled by an exponential distribution.

\section{Concluding remarks}
\label{sec: 8}

We have described a mechanism whereby a gas cloud fragments due to the formation of shocks. When these occur, only a finite fraction of the material in the cloud becomes shocked and undergoes fragmentation. The remainder of the gas remains unaffected, and continues to evolve at a slower rate than the shocked material. This residual material may itself form further shocks as it collapses, or it may become stabilised against further collapse by centrifugal effects. Whatever is the detailed sequence of the collapse process, when non-fragmenting dense cores form, they will do so in an environment of low density gas, which was not able to take the fastest route to gravitational collapse. This reservoir of low density gas can end up feeding the growth of protostars by accretion processes. A theoretical description of this late stage is outside the scope of the theory presented here: we cannot readily estimate how much the mass of a nascent protostar will be increased by accretion processes, nor the effect of the accretion on the form of the initial mass function of stars.

It is instructive to compare our approach with the turbulent fragmentation theory advanced by \cite{Pad+02,Pad+04}. Shock waves play a central role in both theories, but in other respects the theories are very different.
The turbulent fragmentation theory suggests an origin for the Salpeter mass distribution, a power-law distribution for the mass of heavy stars. The power-law in the mass distribution is proposed to result from a power-law relation between relative velocity and separation in the gas cloud, analogous to that which characterises fully-developed turbulence (discussed by \cite{Fri97}). This implies that the shock strength is predicted to increase with the size of the structures in the turbulent gas cloud \citep{Pad+04}.
By way of contrast, in our model the shock strengths approach a universal distribution which is independent of the size scale. Also, we predict that the initial mass of fragments is much smaller than the typical stellar mass, with the protostellar mass determined by subsequent accretion. In the turbulent fragmentation theory the peak of the mass distribution is determined by some adjustable parameters, whereas in our model the fragment mass is determined by radiation transfer considerations.

Our model is consistent with the numerical simulation by \cite{Bat+05} in that the gravitationally bound cores which form protostars are initially much smaller than typical stellar masses. One point of difference is that our model predicts that sheets of dense material formed by shocks will fragment into clouds of material with a size similar to the thickness of the sheet, whereas it is usually reported that simulations show sheets tending to break up into filamentary structures. In our discussion of the gravitational instability we assumed that the local motion of the shocked material is purely rotational. We hypothesise that filamentary structures arise when the local velocity gradient in the shocked layer is anisotropic.

It has been suggested that some of the puzzling features of extrasolar planetary systems, such as the high probability of finding highly eccentric orbits, can be explained by the hypothesis that planets formed by gravitational collapse \citep{Rib+07,Wil+08} rather than by the standard model \citep{Saf69} involving aggregation of dust particles in a circumstellar disc. According to our theory, the size of the smallest fragments produced by gravitational collapse are comparable to the sizes of gas-giant planets, which lends support to the picture of planet formation described by \cite{Wil+08}. In this picture the planets would arise from dense cores which were not able to grow by accretion because of competition from a larger core nearby, which grows to become a protostar.

\end{document}